\documentclass{iau_FM}
\usepackage{graphicx}

\RequirePackage[]{natbib}

\title[Massive binaries] 
{Observational constraints on massive binaries}

\author[Tomer Shenar]   
{Tomer Shenar$^1$
}

\affiliation{$^1$Tel Aviv University, The School of Physics and Astronomy, Tel Aviv 6997801, Israel \\ email: {\tt tshenar@tau.ac.il} }

\pubyear{2024}
\jname{Astronomy in Focus, Focus Meeting 4} 
\editors{Diana M.~Worrall, ed.}
\begin{document}

\maketitle

\begin{abstract}
Binary interactions are commonplace among massive stars, giving rise observed phenomena such as X-ray binaries, stripped stars \& supernovae, and gravitational-wave sources. The multiplicity properties of massive stars thus represent a fundamental observable to calibrate, test, and benchmark models of single-star and binary evolution. In these proceedings, I provide a modern summary of the observed properties of massive binaries across various metallicities, and discuss open problems in the field.
\end{abstract}

\firstsection 
\section{Introduction}

Spectroscopic and interferometric studies of massive stars -- stars initially more massive than $8\,M_\odot$ -- have revealed beyond doubt that the majority of them evolve alongside nearby stellar companions \citep[e.g.,][]{Abt1983, Vanbeveren1998, Sana2012, Sana2013, Sana2014, Moe2017}. With typical physical separations not exceeding a few au,  the majority of massive stars will interact with their companion stars during their lifetime, typically during the rapid expansion phase that follows core hydrogen exhaustion \cite[see, e.g.,][and references therein]{Marchant2023}. These interactions lead to a plethora of observable counterparts that are unique products of binary interaction: post mass-transfer stipped stars and mass accretors \cite[e.g.,][]{Paczynski1967, Goetberg2018, Shenar2020LB1, Bodensteiner2020HR6819, Wang2020, Klement2022}, single degenerate binaries and X-ray binaries \cite[e.g.,][]{Corral-Santana2016, Shenar2022BH, Mahy2022}, magnetic stars \cite[e.g.][]{Ferrario2009, Schneider2018, Shenar2023} and gravitational-wave (GW) sources \cite[e.g.][]{deMink2016, Mandel2022Merging}.

In these proceedings, I provide a brief outline of our current knowledge of the multiplicity statistics of massive stars across various key evolutionary phases and metallicity conditions.  Extensive reviews of the subject can be readily found in recent literature (e.g., \citealt{Marchant2023, Moe2017, Offner2023}). 

\section{Multiplicity statistics: pre-interaction}

The pre-interaction phase is represented by populations of massive stars that are still on the main sequence, which typically appear spectroscopically as O and early B-type stars (B2 and earlier). Multi-epoch spectroscopic surveys of massive stars have thusfar been the most efficient way to study the multiplicity properties of OB binaries, which typically span the orbital period range $
0 \lesssim \log P/{\rm [d]}
 \approx  \lesssim 3$, for which spectroscopy is adequate. Spectroscopic surveys have systematically yielded high observed binary fractions of the order of 30-60\% for O-type stars ($M \gtrsim 15\,M_\odot$), with some variance depending on the environment \citep[e.g.][]{Kobulnicky2014, Sana2012, Sana2013, Britavskiy2023}, dropping to somewhat lower fractions of 20-50\% for B-type stars \citep[e.g.][]{Abt1990, Banyard2022, Bodensteiner2021, Dunstall2015, Villasenor2021}. When correcting for biases, the intrinsic fractions however are found to be comparable and of the order of 50-70\% for both populations.

The distributions of orbital periods $P$ and mass ratios $q = M_2/M_1$ ($M_2 < M_1$) of pre-interaction binaries are typically assumed to follow a power-law distribution in the form $N \propto \left( \log P \right)^\pi\,q^\kappa$, where $\pi$ and $\kappa$ are the power-law indices of the period and mass-ratio distributions, respectively. Derived periods of OB-type binaries follow very closely a flat distribution on $\log P$ ($\pi = 0$), implying that the periods generally tend to be short. For example, after correcting for observational biases, \citet{Almeida2017} report an index of $\pi = -0.1$ for the power-law distribution in the form $\left(\log P\right)^\pi$. Similarly, the mass-ratio distribution is found to be very closely flat of $q$ ($\kappa = 0$). For example, after correcting for biases, \citet{Shenar2022TMBM} report $\kappa = 0.2\pm0.2$ for O-type stars in the LMC.

\section{Dependence on metallicity}

The dependence of multiplicity statistics on metallicity $Z$ can inform models of star and binary formation, and help calibrate evolution models in different environments. Specifically, conditions of multiplicity statistics at sub-solar metallicity, present in galaxies such as the Large and Small Magellanic Clouds (LMC, SMC), are of pivotal importance for understanding electromagnetic and GW transients originating largely or elusively in low-metallicity environments \citep[e.g.][]{Woosley2006, Quimby2011, Fryer2001, Modjaz2008, Mandel2022Merging}. 

Data compiled in \citet{Offner2023} suggests that, for low-mass stars, a trend of increased binary fraction with decreasing metallicity exists. Assuming a linear model on $\log Z$ of the form $f_{\rm bin} = m\,\log Z + b$, the latter study suggests $m = -0.20\pm0.04$ for low-mass stars. For massive stars, obtaining such information is more challenging given the lack of high-quality data and statistics on the multiplicity of low-metallicity OB-type populations. One survey put forth to improve this situation is the Binarity at LOw Metallicity (BLOeM) survey, which will monitor $\approx 1000$ massive stars in the SMC, at a metallicity content of $Z = 1/5\,Z_\odot$ \citep{Shenar2024}. By combining results from spectroscopic surveys of O-type stars in the Milky Way \citep{Sana2012}, LMC ($Z \approx 1/2 Z_\odot$, \citealt{Sana2013}), and early results from the BLOeM survey (Sana et al.\ submitted), no such trend is apparent for O-type stars. Formally, one finds $m = -0.07 \pm 0.15$ for the slope, which is consistent with being metallicity-independent. However, it is also consistent with the relation found by \citet{Offner2023} for low-mass stars, within 1$\sigma$. For B-type stars, combining results from \citet{Banyard2022}, \citet{Dunstall2015}, and early BLOeM results (Villase\~nor et al.\ submitted to A\&A), a regression yields $m = -0.32 \pm 0.14$, which is consistent with results published by \citet{Offner2023} for low-mass stars. This difference between O- and B-type stars may mark a different mode of binary formation for these mass ranges, although this cannot yet be claimed with high certainty given the current  uncertainties on the binary fractions.

\section{Multiplicity statistics: post-interaction}

Unlike the pre-interaction phase, the post-interaction phase has a plethora of observable counterparts. In these proceedings, I addressed only a few important examples in this context. 

\noindent
{\bf Wolf-Rayet stars:} After interaction, the mass donor typically loses substantial amounts of mass, exposing its helium-rich layers. At the upper-mass end, these stars tend to launch powerful radiatively-driven stellar winds and thus spectroscopically appear as Wolf-Ryaet (WR) stars. WR binaries thus offer an attractive laboratory to study the post-interaction phase at the upper-mass end, although they can also form as single stars via intrinsic wind stripping given the high mass of their progenitor stars \citep{Meynet2005}. Generally, the observed binary fraction of WR stars is found to be of the order of 40\%, regardless of the environment \citep{Foellmi2003, Schnurr2008, Neugent2014, Shenar2020, Dsilva2020, Dsilva2022, Dsilva2023,  Schootemeijer2024}. Correction for biases to obtain the intrinsic fractions is more challenging than for the pre-interaction case, since our knowledge of the period and mass-ratio distribution of WR binaries is heavily model-dependent (e.g., \citealt{Langer2020}). Regardless, there is growing evidence that roughly half of the WR stars do not have companions related to their formation. For example, \citet{Schootemeijer2024} conducted a modern high-resolution UVES/VLT spectroscopic monitoring of WR stars in the SMC, but did not reveal companions to any of the WR stars previously classified as single in the period range $P \lesssim 3\,$yr and mass ratio range $0.1 \lesssim q \lesssim 1$. Similar results are reported by \citet{Massey2023} for a class of low-luminosity WR stars in the LMC. Additionally, \citet{Deshmukh2024} conducted an interferometric GRAVITY/VLTI survey of Galactic WR stars, finding similar results, but for wider separations and more extreme mass ratios.  However, the formation of WR stars via wind stripping still poses a challenge for single-star models, and the question remains whether apparently-single WR stars are the products of binary evolution with the companions no longer present (e.g., mergers or disrupted systems).

\vspace{0.1cm}

\noindent
{\bf Stripped stars:}
At lower masses, stars stripped off their outer layers retain a spectroscopic appearance similar to those of OB-type subdwarfs (sdBs, sdOs), since they lack the strong winds which give rise to the Wolf-Rayet phenomenon \citep{Dionne2006, Goetberg2018, Shenar2020, Sander2020}. Such objects are faint in the visual band and, lacking the strong emission lines charactaristic of WR stars, are much more difficult to spot, even though they should be more common than WR stars.  Recently, the first sample of "intermediate mass stripped stars" was uncovered by \citet{Drout2023} and \citet{Goetberg2023}. Given the lower masses of their progenitor stars (initial masses $M_{\rm ini} \lesssim 20\,M_\odot$ in the Milky Way), it is likely that such objects form solely via binary stripping, and hence one may expect that the vast majority of them, if not all, will harbour binary companions. Analysis of follow-up spectroscopy of this sample is currently underway. 

\noindent
{\bf OBe stars:} angular-momentum transfer during mass accretion is thought to efficiently spin up the mass accretors to near-critical rotation \citet{Packet1981}. A well-known class of rapidly-rotating stars are the OBe stars, which are OB-type stars with emission features associated with decretion disks. The last years have yielded overwhelming evidence that the majority, if not all, of OBe stars are the products of binary interaction. Virtually all OBe stars with confirmed companions are clear post-interaction systems, in which the companions are either compact objects or stripped stars of different types \citep[e.g.][]{Wang2018, Wang2020, Klement2022, Shenar2020LB1, Ramachandran2023}. In contrast, no massive OBe star has a confirmed main sequence companion \citep{Bodensteiner2020Be}. Taken together, this strongly points towards a binary origin of OBe stars. However, robust statistical information regarding the binary fraction and the nature of companions to OBe stars (e.g., fraction of black holes neutron stars, white dwarfs, and stripped stars) is still missing. 

\noindent
{\bf Single-degenerate binaries:} after the first core-collapse, the binary system may be disrupted due to the combination of kicks and mass loss \citep[e.g.][]{Podsiadlowski2004}. If the system survives this disruption, one may expect the formation of a single-degenerate binary comprising a massive OB star and a compact object (OB+co). Until recently, such objects were only known where the compact object would be close enough to accrete material from the donor OB star \citep[e.g.][]{Reig2011, Corral-Santana2016}. Such high-mass X-ray binaries (HMXBs) are relatively easy to find via X-ray surveys, but are physically rare compared to the general population of OB+co binaries. Only recently has the latter population of binaries, also referred as dormant single-degenerate binaries, started to be uncovered. Dormant single-denegerate binaries, especially in which the compact object is a black hole, offer a direct route to test and calibrate models involving supernova explosions and kicks \citep[e.g.][]{VignaGomez2024}. The Gaia mission should be able to uncovered tens, perhaps hundreds, of such systems in our Galaxy \citep{Janssens2022, Janssens2023}, although highly stringent quality constraints on the binary solutions so far published by Gaia only allowed low-mass single-degenerate systems to be uncovered \citep{El-Badry2023-GaiaBH1, El-Badry2023_GaiaBH2, Shahaf2023, Panuzzo2024}. Spectroscopic monitoring of massive stars in the Milky Way and LMC revealed two systems with dormant black holes in them \citep{Mahy2022, Shenar2022BH}. The combination of high-precision astrometry and continuous spectroscopic monitoring are expected to increase the sample size of dormant single degenerates by orders of magnitudes in the coming years.

\bibliographystyle{aa}  
\bibliography{papers}

\end{document}